\documentclass{pasa}%
\usepackage{natbib}
\usepackage{soul,color}
\def\url#1{\expandafter\string\csname #1\endcsname}

\def\actaa{Acta. Astronom.} %
\def\aj{AJ}%
\def\araa{ARA\&A}%
\def\apj{ApJ}%
\def\apjl{ApJ}%
\def\aap{A\&A}%
\def\aaps{A\&AS}%
\def\mnras{MNRAS}%
\def\pasp{PASP}%
\title[Extinction Toward the Galactic Bulge]{The Interstellar Extinction Toward the Milky Way Bulge with Planetary Nebulae, Red Clump, and RR Lyrae stars}
\author[Nataf et al.]{David M. Nataf$^1$\thanks{david.nataf@anu.edu.au}\\
\affil{$^1$Research School of Astronomy and Astrophysics, Australian National University, Canberra, ACT 2611, Australia}%
}
\jid{PASA}
\doi{10.1017/pas.\the\year.2015}
\jyear{\the\year}

\begin{document}%
\begin{abstract}
I review the literature covering the issue of interstellar extinction toward the Milky Way bulge, with emphasis placed on findings from planetary nebulae, RR Lyrae, and red clump stars. I also report on observations from HI gas and globular clusters.  I show that there has been substantial progress in this field in recent decades, most particularly from red clump stars. The spatial coverage of extinction maps has increased by a factor $\sim 100 \times$ in the past twenty years, and the total-to-selective extinction ratios reported have shifted by $\sim$20-25\%, indicative of the improved accuracy and separately, of a steeper-than-standard extinction curve. Problems remain in modelling differential extinction, explaining anomalies involving the planetary nebulae, and understanding the difference between bulge extinction coefficients and ``standard" literature values.
\end{abstract}
\maketitle%
\section{INTRODUCTION }
\label{sec:intro}
The measurement of interstellar absorption or scattering of light was first made by \citet{1930LicOB..14..154T}, who showed that the integrated luminosities of star clusters in the Milky Way fell faster than their apparent size\footnote{ \citet{1930LicOB..14..154T} inferred an interstellar extinction of 0.67 photographic magnitudes per kpc. That is, impressively, not spectacularly inconsistent with the modern estimate of 0.7 mag kpc$^{-1}$ in $V$-band  \citet{2006A&A...453..635M}.}. 

It is hard to identify when the first estimates of interstellar extinction toward the bulge were made. However, it is clear that several of the earlier estimates were toward globular clusters in the direction of the bulge. Colour excesses were estimated toward 68 globular clusters by \citet{1936ApJ....84..132S}, including several bulge globular clusters such as NGC 6440 and NGC 6441.  \citet{1965ApJ...141...43A} measured $E(B-V)=0.46 \pm 0.03$ toward NGC 6522, which is in the same direction as Baade's window \citep{1951POMic..10....7B}. 

Reddening measurements toward the field were sparser. \citet{1971AJ.....76.1082V,1972PASP...84..306V} measured $E(B-V)=0.45$ toward field stars in the vicinity of NGC 6522, mitigating a potential error, that NGC 6522 might be at a different distance than the stellar population of Baade's window and thus perhaps having a different integrated reddening. Another early measurement is that of $E(B-V)=0.25 \pm 0.05$ \citep{1974AJ.....79..603V} toward $(l,b)=(0,-8)$, now known as Plaut's window \citep{1973A&AS...12..351P}. These sightlines, chosen for study due to their relatively low extinction, have been important in the historical development of bulge studies, as they were often selected for more detailed investigations \citep{1988AJ.....96..884T,1995AJ....110.2788T,1998A&A...338..581S,2007AJ....134.1432V,2011ApJ...732..108J}. 

The modern era in investigations of extinction toward the bulge, ultimately resulting from the development and implementation of CCD technology, began with the high-resolution extinction map of Baade's window by \citet{1996ApJ...460L..37S}. Stars along the red clump -- the horizontal branch of an old, metal-rich stellar population -- were used as standard candles to measure the shift in colour and magnitude as a function of direction. The advantages of this method are that red clump stars are bright, numerous, occupy a relatively narrow position in the colour-magnitude diagrams, and can in principle have calibrated zero points. This method has been extended and deployed by several other groups to constrain each of the reddening and extinction toward ever-larger fractions of the bulge \citep{2004MNRAS.349..193S,2008A&A...491..781C,2009ApJ...696.1407N,2010A&A...515A..49R,2012A&A...543A..13G,2013ApJ...769...88N,2013MNRAS.435.1874W,2016MNRAS.456.2692N} and toward other stellar systems in the local group \citep{1998ApJ...503L.131S,2000ASPC..203..203P,2005A&A...430..421S,2009MNRAS.397L..26C,2010ApJ...721..329C,2011ApJ...727...55M}. The same Galactic bulge data sets have often been used to trace the reddening with RR Lyrae stars \citep{1999ApJ...521..206S,2008AJ....135..631K,2012ApJ...750..169P,2013ApJ...776L..19D,2015ApJ...811..113P}, yielding mostly complementary results. 

There have also been investigations of the reddening and extinction toward the inner Milky Way with planetary nebulae. The assumption informing that research is that the intrinsic ratio of fluxes from separate lines and even the radio continuum can be robustly predicted from theory \citep{1971MNRAS.153..471B}. The ``reddening", in this case an excess in the ratio of flux of one line with respect to another, can then be converted to extinction, necessary to solving for the intrinsic luminosity for the planetary nebulae as well as a location within the Galaxy (via the distance), both critical to using the planetary nebulae to discuss Galactic evolution. This research program has had similar findings to that of tracing the reddening from red clump and RR Lyrae stars, typically finding that the extinction is high and non-standard in its wavelength-dependence \citep{1992A&A...266..486S,1992A&AS...95..337T,1993IAUS..153..337W,2004MNRAS.353..796R,2012IAUS..283..380H} with some objections \citep{2013A&A...550A..35P}. 

Many readers would see the development of these two literatures and assume that this has been used for confirmation or negation, but this was not the case. The literature of investigating the reddening toward the bulge with red clump and RR Lyrae stars has developed completely independently and in parallel to that of planetary nebulae. I find no references to the research findings from planetary nebulae in the recent, high-impact papers from \citet{2012A&A...543A..13G} and \citet{2013ApJ...769...88N}. Similarly and conversely,  \citet{2013A&A...550A..35P} does not reference any of the literature on reddening toward the bulge involving red clump and RR Lyrae stars. That is  a failing of these two branches of the astronomy literature, which I hope will be partially rectified by this review where both areas are discussed. In principal the systematics and wavelength-dependence of these two methods are distinct, so there is great potential for synergy between the two areas. At the same time, they both benefit from the relatively high reddening toward the bulge, which reduces the relative weight and concern of zero-point uncertainties. 

The structure of  this review is as follows. In Section \ref{sec:PNe} I review the literature of Galactic bulge reddening estimates from planetary nebulae. In Section \ref{sec:H1emission}, I briefly discuss measurements of the extinction curve from H1 emissions by means other than planetary nebulae. In Section \ref{sec:RCRRLyrae} I review the findings by means of RR Lyrae and red clump stars. In Section \ref{sec:Globulars}, I briefly review the measurements from globular clusters.The discussion and conclusion are presented in Section \ref{sec:Conclusion}.

\section{Measuring Extinction with Planetary Nebulae}
\label{sec:PNe}
There are two widely-used methods to measure extinction from measurements of planetary nebulae. I briefly introduce these prior to presenting and discussing literature results.

\subsection{The Balmer Decrement}
The first method is that of the Balmer decrement. Here, the observed ratio of intensities of the $3 \rightarrow 2$ ($H_{\alpha}$, 6,563 \AA) and $4 \rightarrow 2$ ($H_{\beta}$, 4,861 \AA) transitions of the hydrogen atom are compared to their intrinsic intensity ratio, so as to yield a relative extinction. The intrinsic intensity ratio, $H_{\alpha,0}:H_{\beta,0} \approx 2.85$, has negligible dependence on temperature and density \citep{1971MNRAS.153..471B}. Given that interstellar extinction is a decreasing function of wavelength for $\lambda \gtrsim 2,500\, \AA$, the measured intensity ratio for a reddened planetary nebulae will always be $H_{\alpha}:H_{\beta} > 2.85$, where we interchangeably use $H_{\alpha }$ and $H_{\beta}$ to refer to each of the name of the atomic transition, the photon emission feature, and the measured flux of the line, as is customary in the literature. One can thus derive an extinction of the $H_{\beta}$ by measuring the intensity ratio $H_{\alpha}:H_{\beta}$, and assuming an interstellar extinction curve, as follows:
\begin{equation}
A_{4861} = -2.5 \log_{10} \{ H_{\beta}/H_{\beta,0}  \},
\end{equation}
\begin{equation}
A_{6563} = -2.5 \log_{10} \{ H_{\alpha}/H_{\alpha,0}  \},
\end{equation}
and thus:
%\begin{equation}
%\begin{split}
%A_{4861} = \frac{A_{4861}}{A_{4861} - A_{6563}} {\times}  \\
%\biggl(   -2.5 log_{10} \{ H_{\beta}/H_{\beta,0} \} +  2.5 \log_{10} \{ H_{\alpha}/H_{\alpha,0} \}     \biggl) = \\
%  \frac{A_{4861}}{A_{4861} - A_{6563}} {\times} 2.5 \log_{10} \{  \frac{H_{\alpha}/H_{\beta}}{2.85}    \},
%\end{split}
%\end{equation}
%and therefore:
\begin{equation}
C_{\rm{opt}} \equiv A_{4861} = 2.5 \frac{A_{4861}}{A_{4861} - A_{6563}}   \log_{10} \{  \frac{H_{\alpha}/H_{\beta}}{2.85}  \},
\label{EQ:3}
\end{equation}
where $C_{\rm{opt}}$ is an arbitrary notation for $A_{4861}$ that is widely used in the literature, with $C_{bd}$ also sometimes used.  The value of $A_{4861}/(A_{4861} - A_{6563})$ is a function of the interstellar extinction curve, we list some specific cases in Table \ref{table:1}. It is clear that the value of $A_{4861}/(A_{4861} - A_{6563})$ is a somewhat sensitive function of the both the extinction curve parameter $R_{V}$ and which reference is used to provide the parameterization. 

\begin{table}
\centering
\begin{tabular}{|c|c|r|}
	\hline \hline
$\frac{A_{4861}}{(A_{4861} - A_{6563})}$ & $R_{V}$ & Reference \\
	\hline \hline \hline
3.07 & 3.17 & \citet{1979MNRAS.187..785S} \\
3.09 & 3.2 & \citet{1979MNRAS.187P..73S}  \\
3.22 & 3.1 & Savage \& Mathis (1979) \\
2.68 & 2.1 & \citet{1989ApJ...345..245C} \\
3.36 & 3.1 & \citet{1989ApJ...345..245C} \\
3.92 & 4.1 & \citet{1989ApJ...345..245C} \\
2.17 & 2.1 & \citet{1999PASP..111...63F}  \\
2.87 & 3.1 & \citet{1999PASP..111...63F} \\
3.52 & 4.1 & \citet{1999PASP..111...63F} \\
2.83 & 3.0 & \citet{2007ApJ...663..320F} \\
3.03 & 3.1 & \citet{2016arXiv160203928S} \\
\hline
\end{tabular}
\caption{The value of the Balmer decrement coefficient  ${A_{4861}}/{(A_{4861} - A_{6563}})$ as a function of the interstellar extinction curve parameter $R_{V}$ and the chosen bibliographic reference, for several representative values. Even if one fixes $R_{V}$ to the ``standard" value, there remains  a margin of manoeuvre of $\sim$17\%. }
\label{table:1}
\end{table}

In principle, the method outlined here could be extended to include the other Balmer lines (e.g. \citealt{2012MNRAS.419.1402G}), in practice the signal-to-noise for lines such as $H_{\gamma}$ is often too low to be of use, and thus it has not been widely used in the literature of Galactic bulge planetary nebulae. 

\subsection{Radio Continuum}
From \citet{1984ASSL..107.....P}, see also \citet{2004MNRAS.353..796R}, the intrinsic radio  to $H_{\beta}$ flux ratio is given by:
\begin{equation}
S_{\nu}/F_{H_{\beta,0}} = 2.51 \times 10^{7} T_{e}^{0.53} \nu^{-0.1} Y \rm{(Jy\,mW^{-1}\,m^2)},
\end{equation}
where $S_{\nu}$ denotes the flux in the radio, $T_{e}$ is the electron temperature in Kelvin, $\nu$ is the radio frequency in Ghz,  and $Y$ is a factor incorporating the ionized helium-to-hydrogen ratio. The latter is not to be confused with the initial mass fraction of stars composed of helium, which is also conventionally denoted $Y$. It is conventionally assumed that there is no opacity in the radio, and thus $S_{\nu,0} = S_{\nu}$, and thus we simply write $S_{\nu}$. For standard values of these parameters $T_{e} = 10^4$ K, $\nu = 5$ Ghz, and $Y=1.1$, and converting to mJy, one gets:
\begin{equation}
F_{H_{\beta,0}} = 3.23 \times 10^{-13} S_{\nu}  \rm{(mJy^{-1}\,mW\,m^{-2})}. 
\end{equation}
This estimated intrinsic flux in the $H_{\beta}$ line can be compared to the observed flux, yielding an estimate of the extinction :
\begin{equation}
C_{\rm{rad}} \equiv  -2.5 \log_{10} \{ H_{\beta}/ 3.23 \times 10^{-13} S_{\nu}   \}.
\label{EQ:6}
\end{equation}

Within the literature, the resulting extinction estimates from Equation \ref{EQ:3} and Equation \ref{EQ:6} are compared, with discrepancies frequently associated with variations in the optical extinction curve, as per Table \ref{table:1}. \citet{2013A&A...550A..35P}  have suggested that there may be systematic errors in $C_{\rm{rad}}$ due to radio opacity within the planetary nebulae themselves. 

\subsection{Literature Results: Steeper-than-Standard Extinction Suggested but not Confirmed}
\citet{1992A&A...266..486S} compared measurements of $C_{\rm{opt}}$ and $C_{\rm{rad}}$ for $\sim$130 Galactic planetary nebulae. They found a mean offset of 20\% in their sample between the two estimates, which they argued was due to a steeper extinction curve with $R_{V}$ significantly lower than 3. From Table \ref{table:1} of this review, a 20\% reduction in $C_{\rm{opt}}$ could be achieved by a shift in the extinction curve of $\Delta R_{V} \approx 0.8$, though the exact value depends on the preferred formalism. \citet{1992A&A...266..486S} don't conclude that the extinction is steeper toward the bulge, as latter works did. Rather, they say that ``standard" extinction curves as defined by the literature merely describe the properties of the interstellar medium within $\sim$2 kpc of the Sun. The interstellar medium in other areas of the Galaxy, for example the regions between spiral arms, are characterized by smaller values of $R_{V}$.

Concurrently, \citet{1992A&AS...95..337T} studied a sample of 900 Galactic planetary nebulae. They found that estimates of extinction from the Balmer decrement and the radio continuum were consistent for planetary nebulae with low extinction, but that at high extinction, the Balmer decrement overestimated the extinction. Their suggested explanation was that the faintest planetary nebulae also had underestimated radio fluxes, due to measurement errors. 

A third study from that period, that of \citet{1992A&AS...94..399C}, compiled a list of H$\beta$, HeII $\lambda$4686 fluxes, 5 Ghz radio flux densities, and the Balmer decrement for 778 Galactic planetary nebulae. They also observed that extinction determinations from the Balmer decrement and the radio continuum were consistent for low-extinction planetary nebulae, but not for high-extinction planetary nebulae.  No attempt was made to attribute the offset to Galactic environment. Rather,  \citet{1992A&AS...94..399C} speculate that regions of high extinction might have a different total-to-selective extinction ratio. 

\citet{1994A&A...289..261P} followed up the controversy by obtaining more reliable measurements of the radio fluxes of 20 planetary nebulae. The measurements of the flux were made at 6cm wavelengths and 3.6cm wavelengths. Roughly half their sample were of bulge planetary nebulae. \citet{1994A&A...289..261P} confirmed that there were measurement errors in the radio flux for the faintest planetary nebulae. However, even with their improved measurement, an offset remained between the extinction estimates from the Balmer decrement and the radio continuum. They discuss two possible resolutions: either the extinction toward the faintest planetary nebulae is in fact steeper-than-standard, or the faintest planetary nebulae also have internal opacity in the radio. 

\citet{2004MNRAS.353..796R} studied 70 planetary nebulae, for which they obtained values of the angular diameter, flux, as well as extinction from narrow-band filter photometry centred on $H_{\alpha}$ and [OIII], where the latter line is located at 5,007 \AA. By means of this more robust dataset, they suggest a mean extinction curve toward the bulge of $<R_{V}>=2.0$, assuming the parameterization of \citet{1989ApJ...345..245C}. \citet{2012IAUS..283..380H} followed up the issue by obtaining more precise, more accurate radio measurements, obtained with the Australian Telescope Compact Array. They found a distribution in extinction curve parameter, spanning the range $0.84 \leq R_{V} \leq 2.85$, with the same mean value $<R_{V}>=2.0$ as \citet{2004MNRAS.353..796R}.

%%%%%%%% pottasch & Bernard salas mentioned here
 \citet{2013A&A...550A..35P} have rejected this hypothesis of low $R_{V}$ by means of a novel, alternative method. Concerned by the possibility of the planetary nebulae's internal opacity in the radio continuum, they obtained measurements of hydrogen lines in the mid-infrared region with data from the \textit{Spitzer} Space Telescope for 16 planetary nebulae. Specifically,  they measured the flux of a line at 7.46$\mu$m that is a blend of lines from the $6 \rightarrow 5$ (Pfund ${\alpha}$) atomic transitions with the $8 \rightarrow 6$ and $17 \rightarrow 8$ atomic transitions; as well as a line at $12.37\,{\mu}$m that is a blend of $7 \rightarrow 6$  (Humphreys ${\alpha}$) atomic transition with the  $11 \rightarrow 8$ transition. These blends are corrected for. Together, these two line strengths can be used to predict the unextincted H$\beta$ flux, and the difference with the measured H$\beta$ flux is denoted $C_{ir}$. The advantage of $C_{ir}$ is that it is independent of assumptions as to the internal radio opacity of planetary nebulae. 

This allows a different, arguably more robust comparison to $C_{\rm{opt}}$. \citet{2013A&A...550A..35P}  use the theoretical results from Table 6 of \citet{1987MNRAS.224..801H}, that  $\rm{Pfund}\,\alpha:H_{\beta,0} \approx 2.45 \times 10^{-3}$ and  $\rm{Humphreys}\,\alpha:H_{\beta,0} \approx 9.27 \times 10^{-3}$, to derive a value of $H_{\beta}$ that is independent of extinction, given the assumption that extinction uncertainties in the mid-infrared are negligible. Indeed, recent results from both \citet{2009ApJ...707..510Z} and \citet{2009ApJ...696.1407N} measure $A_{[8.0\mu]}/A_{Ks} \approx 0.40$, so  $A_{[8.0\mu]}/A_{Ks} \lesssim 0.03A_{V}$  \citep{2008ApJ...680.1174N}.  \citet{2013A&A...550A..35P} use the extinction corrections from \citet{2006ApJ...637..774C}.   

\begin{figure}
\begin{center}
\includegraphics[totalheight=0.40\textheight, angle =90]{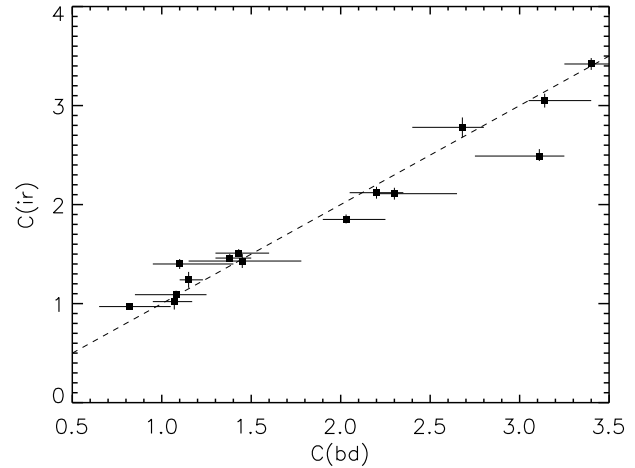}
\caption{This is Figure 1 from \citet{2013A&A...550A..35P}. Shown is the comparison between the extinction of the H$\beta$ line predicted by two different methods. The dashed line denotes where the points would lie if the measurements and predictions were both perfect.}
\label{figPottasch1}
\end{center}
\end{figure}

In their Figure 1 (which is shown here as Figure \ref{figPottasch1}), \citet{2013A&A...550A..35P} plot the scatter of $C_{IR}$ vs $C_{\rm{opt}}$  and state that they appear similar to the prediction for $R_{V}=3.1$. The conclusion of \citet{2013A&A...550A..35P} is that the extinction toward the bulge is well-described by the $R_{V}=3.1$ extinction curve from \citet{1979ARA&A..17...73S}. That may be -- however it does appear problematic that the unity line in Figure \ref{figPottasch1} is clearly not the best-fit line, which would have a shallower slope and a non-zero intercept. It is also of interest that the results of \citet{2013A&A...550A..35P} suggest that the conversion of radio flux to H$\beta$ flux (Equations 4 and 5) fails in the specific case of Galactic bulge planetary nebulae, even as it does not lead to discrepancies for planetary nebulae within 2 kpc of the Sun. 

%would imply by arguments mentioned above that the planetary nebulae in the bulge are distinct from planetary nebulae elsewhere, that they have higher internal opacity in the radio continuum. \citet{2013A&A...550A..35P}  provide no explanation for this consequence of their conclusions. 

\subsection{Ultraviolet Extinction Toward the Bulge with Planetary Nebulae}
There is no vast literature on the ultraviolet extinction toward the Galactic bulge, due to the obvious issue that by being so large it becomes harder to measure.  The one result discussed in this review is that of \citet{1993IAUS..153..337W}. Unfortunately, it is a conference proceeding that does not appear to have been followed up by a detailed publication. The sample and methodology are not discussed in detail. \citet{1993IAUS..153..337W} studied 42 Galactic bulge planetary nebulae, for which they measured $H_{\alpha}$, $H_{\beta}$, $H_{\gamma}$, as well as He II $\lambda$1640 in 4 of their 42 planetary nebulae. They reported measuring a mean optical reddening law of $R_{V}=2.29$. In the ultraviolet, 2 of their planetary nebulae have extinctions $\sim 25$\% higher than expected, and the other 2 have extinctions that are $\sim$80\% higher than expected.

\section{Extinction Toward the Galactic Centre with Hydrogen Line Emission}
\label{sec:H1emission}
\citet{2011ApJ...737...73F} measure the emission of 21 hydrogen lines (wavelength range $1 \leq \lambda/{\mu}m \leq 19$)   from a minispiral gas cloud near the Galactic centre. Among these lines are $5 \rightarrow 3$ (Paschen $\beta$, 1.282 ${\mu}$m), $10 \rightarrow 4$ (Brackett $\zeta$, 1.736 ${\mu}$m), and $7 \rightarrow 4$ (Brackett $\gamma$, 2.166 ${\mu}$m). These are the three lowest-wavelength lines, they also nearly correspond to the standard near-infrared photometric filters $J$, $H$, and $K$. The data from these 21 lines was combined with a map of 2cm emission from the Very Large Array to map the extinction. 

For five lines with wavelength of 2.728 $\mu$m or less, \citet{2011ApJ...737...73F} derive a form for the extinction of A$_{\lambda} \approx \lambda^{-2.11\pm0.06}$, which is steeper than standard, but consistent with other determinations for Galactic centre. For wavelengths between 3.7 and 8.0 $\mu$m, they derive A$_{\lambda} \approx \lambda^{-0.47\pm0.29}$, which is significantly \textit{grayer} than the extinction in the near-infrared, which they showed to be a substantial theoretical difficulty. 

With their maps, \citet{2011ApJ...737...73F} derive extinctions toward Sgr A* of $A_{H}=4.21 \pm 0.10$, $A_{K_{s}}=2.42 \pm 0.10$, and $A_{L'}=1.09 \pm 0.13$, in the NIRC filter system of the Keck Telescope\footnote{\url{http://www2.keck.hawaii.edu/inst/nirc2/filters.html}}, which is relevant to Galactic centre studies. 

\section{Extinction Studies of the Galactic Bulge with RR Lyrae and Red Clump Stars}
\label{sec:RCRRLyrae}
RR Lyrae and red clump stars are almost certainly the two most frequently used tracers of the extinction toward the bulge. Both are sensible tracers given their standardizable colours, magnitudes, relatively high number counts, and relatively high luminosity ($M_{V} \approx +0.50$) making them observable even to comparatively large distances and extinctions.

\subsection{The 1990s-era RR Lyrae and Red Clump Colour Controversy}
The developments of expanding hard drives, CCDs, and dedicated telescopes enabled larger and more uniform photometric surveys such as the \textit{Optical Gravitational Lensing Experiment} (OGLE) \citep{1998AcA....48..113U}, thus allowing more detailed questions to be probed than previously possible. 

One such effort was the analysis of \citet{1999ApJ...521..206S}, which combined photometry from various surveys to obtain the $(V-I)_{0}$ and $(V-K)_{0}$ of RR Lyrae toward Baade's window. They found that the dereddened colours were standard in $(V-K)_{0}$, while being $\sim 0.17$ mag too red in $(V-I)_{0}$. There was no offset in a comparison sample they had of more nearby RR Lyrae. With the information then available, they argued that the anomalous colours were due to an offset in $\alpha$-element abundance. We now know that bulge stars with [Fe/H] $\approx -1.0$, corresponding to RR Lyrae, have identical or nearly identical [$\alpha$/Fe] abundance ratios of local thick disk and halo stars \citep{2013A&A...549A.147B,2013MNRAS.430..836N}, corresponding to local RR Lyrae. The suggestion of \citet{1999ApJ...521..206S} is thus no longer a viable solution to the phenomenon they identified.

A similar offset was measured by \citet{1998ApJ...494L.219P} and \citet{1998AcA....48..405P}. They found that the $(V-I)_{0}$ colour of bulge red clump stars was $\sim$0.20 mag redder than that expected from a calibration to the \textit{Hipparcos} sample of red clump stars. The same offset as \citet{1999ApJ...521..206S} is measured, toward the same sightline, assuming the same extinction curve coefficient of $A_{V}/E(V-I) \approx 2.5$, but using different stellar tracers.  \citet{1998ApJ...494L.219P}  attribute the colour offset in red clump stars as being due to higher metallicity for Baade's window bulge stars than solar neighbourhood stars. Subsequent studies of the properties of the red clump have since shown that a 0.20 mag offset in $(V-I)_{0}$ colour would require a metallicity shift of ${\Delta}$[Fe/H] $\approx 0.75$ dex \citep{2001MNRAS.323..109G,2014MNRAS.442.2075N}, which is completely ruled out by bulge spectroscopic data (e.g. \citealt{2008A&A...486..177Z,2011ApJ...732..108J,2013MNRAS.430..836N}). 

The actual solution to the discrepancy, now understood and demonstrated further in this review, is that \citet{1999ApJ...521..206S} and \citet{1998ApJ...494L.219P} both  underestimated reddening $E(V-I)$ due to their assumption of $A_{V}/E(V-I) \approx 2.5$. This extinction coefficient has since been measured to be $A_{V}/E(V-I) \approx 2.2$ \citep{2013ApJ...769...88N} in the mean, reaching values as low as  $A_{V}/E(V-I) \approx 2.0$ \citep{2004MNRAS.349..193S} . Thus, for fixed $A_{V}$,  an overestimated  $A_{V}/E(V-I)$ leads to an underestimated $E(V-I)$ and thus an overestimated $(V-I)_{0}$. Though one could have technically deduced this discrepancy as being due to non-standard extinction, at the time of these studies there were still some unresolved sources of systematic errors, such as the orientation of the Galactic bar and the detailed chemistry of bulge stars. These are ruled out as explanations today, but they were viable scientific hypotheses at the time. 

\subsection{First-Generation Wide-Field Optical Extinction Maps of the Galactic Bulge}
The OGLE and \textit{MAssive Compact Halo Object} (MACHO) surveys \citep{1999PASP..111.1539A} were the first wide-field photometric surveys of the Galactic bulge, and among the results were the first large extinction maps and investigations of the extinction curve toward the bulge. 

\begin{figure}
\begin{center}
\includegraphics[totalheight=0.40\textheight]{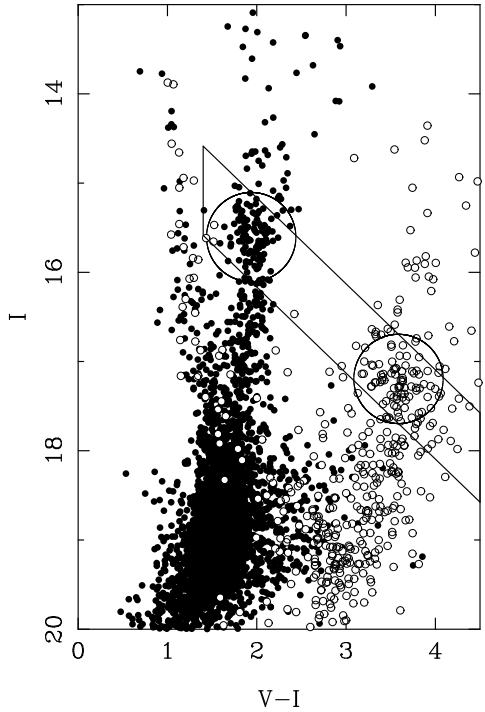}
\caption{This is Figure 5 from \citet{2004MNRAS.349..193S}. Shown is the colour-magnitude diagram toward two sightlines near $(l,b)=(-0.380,-3.155)$, making clear the effect of interstellar extinction and reddening. The red clump shifted to a redder and fainter position.}
\label{figSumi5}
\end{center}
\end{figure}

The first investigations were those of $VI$ photometry in OGLE by  \citet{1996ApJ...460L..37S} and \citet{1996ApJ...464..233W}. The colours and magnitudes of the red clump were compared to their dereddened colours and magnitudes to infer the extinction and extinction curve.  The red clump method can be visualized in Figure \ref{figSumi5}, which is Figure 5 from \citet{2004MNRAS.349..193S}. Evidence was found that the extinction coefficient $A_{V}/E(V-I)$ could differ by as much as 0.37 mag mag$^{-1}$ between different bulge sightlines, but the extent of the variation was degenerate with the spatial properties of the Galactic bar, which were not precisely and convincingly measured at the time. The issue was partially resolved by  \citet{2003ApJ...590..284U}, who used greater spatial coverage to demonstrate that the extinction curve varies even on small scales, over which the Galactic bar will contribute negligible viewing effects. \citet{2004MNRAS.349..193S} confirmed the findings by measuring the reddening and estimating the total-to-selective extinction ratio over the entirity of the OGLE-II Galactic bulge survey \citep{2002AcA....52..217U}, and measured a mean value of $A_{I}/E(V-I)=0.964$, substantially lower than the canonical value of $A_{I}/E(V-I) = 1.48$ \citep{1989ApJ...345..245C}  or $A_{I}/E(V-I) = 1.33$ \citep{1999PASP..111...63F}. 

A contending finding is that of \citet{2008AJ....135..631K}, who used MACHO $VR$ measurements of RR Lyrae stars to study the reddening toward the bulge. The mean total-to-selective extinction ratio measured was $A_{R}/E(V-R) = 3.3 \pm 0.2$ where the fit is obtained using the OLS bisector method of \citet{1992ApJ...397...55F}. In comparison, the prediction from a standard extinction curve is $A_{R}/E(V-R) = 3.34$ \citep{1999PASP..111...63F}. The disagreement between the results of \citet{2008AJ....135..631K}, versus the results of \citet{2003ApJ...590..284U}, \citet{2004MNRAS.349..193S}  and subsequent OGLE results among others is not explained at this time. It may be that the extinction curve differences are more pronounced in the range $ 6500 \lesssim \lambda/\AA \lesssim 8000$ than in the range $ 5500 \lesssim \lambda/\AA \lesssim 6500$. It could also be a selection effect due to the different spatial coverage. The cause of these different results has not been identified at this time. 

\subsection{Extinction Curve Anomalies in the Infrared}
The investigations of \citet{2006ApJ...638..839N}, \citet{2008ApJ...680.1174N}, and \citet{2009ApJ...696.1407N} used photometry of red clump stars to conclusively demonstrate that the near-infrared extinction curve toward the inner Milky Way is non-standard. This is a more surprising result than that of extinction curve variations in the optical, as the works of \citet{1989ApJ...345..245C} and \citet{1999PASP..111...63F} predict a universal extinction curve in the infrared independent of variations in $R_{V}$. 

 \citet{2006ApJ...638..839N} reduced photometry from the IRSF telescope toward $|l| \lesssim 2.0$ and $|b|  \lesssim 1.0$. \citet{2008ApJ...680.1174N} added $V$-band photometry from OGLE to the analysis, reporting a mean value of $A_{V}/A_{Ks} \approx 16$, substantially higher than the standard value of $\sim$9. Finally \citet{2009ApJ...696.1407N} added near-IR calibration from 2MASS \citep{2006AJ....131.1163S} and mid-infrared photometry from Spitzer/IRAC GLIMPSE survey \citep{2003PASP..115..953B}. \citet{2009ApJ...696.1407N} measured the mean extinction coefficients $A_{J}:A_{H}:A_{Ks} :A_{[3.6]}:A_{[4.5]}:A_{[5.8]}:A_{[8.0]} = 3.02:1.73:1:0.50:0.39:0.36:0.43$. The near-infrared extinction toward the Galactic bulge is well-fit by a power-law $A_{\lambda} \propto \lambda^{-2.0}$. For wavelengths longer than 2.2 $\mu$m ($K_{s}$-band), the extinction is greyer than expected from a simple extrapolation, a feature also measured by others \citep{2005ApJ...619..931I,2009ApJ...707..510Z,2009ApJ...707...89G} and predicted by several dust models \citep{2001ApJ...548..296W,2004ApJ...611L.109D,2006A&A...445..167V}. 
 
 \citet{2009ApJ...696.1407N}  find evidence for variations in the infrared extinction curve, but it is not significant. However, though the extinction curve is not confirmed to vary within their observational window, their results differ from other literature results toward other regions of the sky. For example,  \citet{2009ApJ...696.1407N} measure a mean value of $A_{K}/E(H-K_{s}) = 1.44 \pm 0.01$, significantly different from the value of $A_{K}/E(H-K_{s}) = 1.82$ reported by \citet{2005ApJ...619..931I}, by means of a similar methodology and data. 
 
 \subsection{Extinction Curve Anomalies Measured with Hubble Space Telescope Photometry}
\citet{2010A&A...515A..49R} investigated the extinction curve toward a bulge window using photometry from the \textit{Hubble Space Telescope (HST)} in the bandpasses $F435W$ and $F625W$, which roughly correspond to Johnson-Cousins $B$ and $R$, respectively. As their photometry is measured with HST, they have completely independent systematics, for example for issues relating to photometric zero points. 

They report high extinction, with an average measurement of $|A_{F625W}|=4$, with significant variations over their total field of $6.6' \times 6.6'$. Their mean extinction coefficient is $A_{F625W}/(A_{F435W} - A_{F625W}) = 1.25$, corresponding to an $R_{V} = 1.97, 2.46$ depending on whether or not one uses the parameterizations of \citet{1989ApJ...345..245C} or \citet{1999PASP..111...63F}. 

This can be regarded as among the most definitive demonstrations of anomalous extinction toward the bulge. On the other hand, it is a small window, and as pointed out by \citet{2016MNRAS.456.2692N}, it happens to be toward a field where all indicators agree that the extinction curve is exceptionally steep. 
 
% Thus, it is established that the extinction in the wavelength range of J, H, and KS is well fitted by a power law of steep decrease A? ? ??2.0 toward the GC. I
% The extinction curve in the near-IR is well-

 % and measured mean extinction coefficients of $A_{J}:A_{H}:A_{Ks}=3.01:1.75:1$, with small variations of $A_{Ks}/(A_{H}-A_{Ks})$ across their survey. 

\subsection{Near-Infrared Reddening Maps of the Galactic Bulge from the VVV Survey}
The \textit{VISTA Variables in the Via Lactea} $(VVV)$ survey \citep{2012A&A...537A.107S} yielded high-resolution, deep, near-infrared imaging over 526 square degrees of the Galactic bulge. For the first time, the photometrically-discernible properties of the bulge could be discerned on a ``global" scale, rather than via spot duty from a few specifically targeted fields. Among the most successful data products is the global reddening map, which is to be expected given that it is a prerequisite to most other bulge science one can do with $VVV$. 

The methodology was developed by \citet{2011A&A...534A...3G}, who measured the $(J-K_{s})$ colours of the red clump over the region  $0.2 < l < 1.7$,   $-8 < b < -0.4$. Among the first results identified by  \citet{2011A&A...534A...3G} were consistency between resulting photometric metallicity estimates and published spectroscopic results \citep{2008A&A...486..177Z,2011ApJ...732..108J}, as well as the discernibility of the double-peaked luminosity toward high-latitude fields that is due to the peanut/X-shape of the Milky Way bulge \citep{2010ApJ...721L..28N,2010ApJ...724.1491M}.

\begin{figure}
\begin{center}
\includegraphics[totalheight=0.27\textheight]{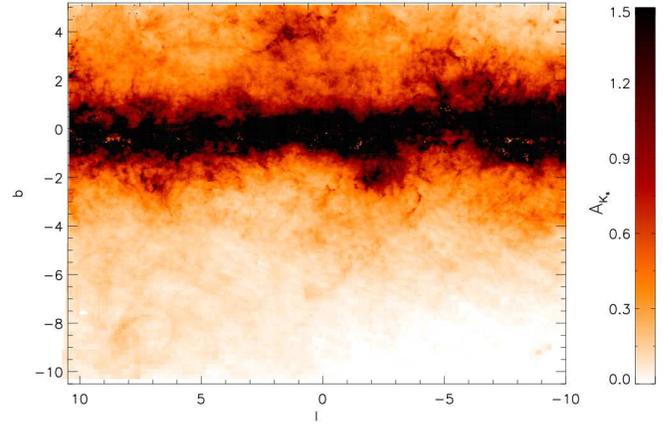}
\caption{This is Figure 3 of \citet{2012A&A...543A..13G}, It is the distribution of $A_{Ks}$ toward the Galactic bulge, as measured in $VVV$ infrared photometry. The scale saturates for $A_{Ks} \gtrsim 1.5$, covering the inner regions, for which the reader is referred to Figure \ref{figGonz6} below.}
\label{figGonz3}
\end{center}
\end{figure}

\begin{figure}
\begin{center}
\includegraphics[totalheight=0.23\textheight]{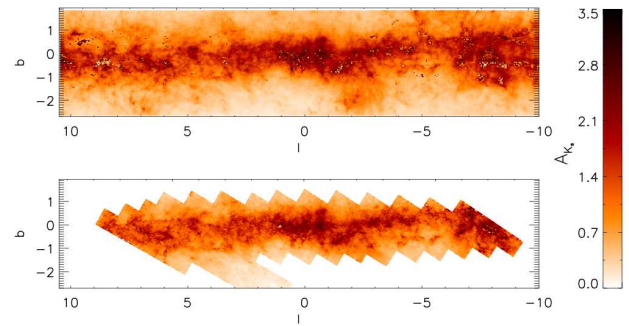}
\caption{This is Figure 6 of \citet{2012A&A...543A..13G}, It is the distribution of $A_{Ks}$ as in Figure \ref{figGonz6}, but zoomed in toward the inner regions to show the variation in extinction there. }
\label{figGonz6}
\end{center}
\end{figure}

\citet{2012A&A...543A..13G}  followed up with a reddening  map over the coordinate range $-10.0 < l < 10.4$,   $-10.3 < b < 5.1$, so $\sim$315 square degrees, broader coverage than any other bulge reddening map released either before or since.  We display Figures 3 and 6 from \citet{2012A&A...543A..13G}, as Figures \ref{figGonz3} and \ref{figGonz6} here. Figure \ref{figGonz3}  is a surface map of the extinction toward the bulge in $A_{Ks}$. Values as high as $A_{Ks} = 3.5$ are measured, corresponding to $A_{V} \approx 50$ \citep{2009ApJ...696.1407N,2016MNRAS.456.2692N}. Figure \ref{figGonz6} is the same, but zoomed into regions of higher extinction closer to the plane to better emphasize the contrast. That kind of extinction is far too high to be measured in the optical, demonstrating the need for both infrared photometry of the inner bulge and the corresponding infrared extinction maps. 

The analysis of \citet{2012A&A...543A..13G}  has since been validated by \citet{2013MNRAS.435.1874W}, who report agreement at the level of ${\Delta}E(J-K_{s}) \approx 0.01$ mag. \citet{2014A&A...569A.103R} also validated the zero-points of the reddening map of  \citet{2012A&A...543A..13G} by comparing photometric and spectroscopic temperatures, measuring an offset of ${\Delta}E(J-K_{s})=0.006 \pm 0.026$ -- consistent with zero. The extinction maps of \citet{2012A&A...543A..13G}, along with metallicity maps from  \citet{2011A&A...534A...3G}, are available for download online at the BEAM calculator's webpage\footnote{\url{http://mill.astro.puc.cl/BEAM/calculator.php}}. 

\subsection{Optical Reddening and Extinction Maps of the Galactic Bulge}
The most recent, up-to-date, and broadest-in-coverage optical extinction maps of the Galactic bulge are those of \citet{2013ApJ...769...88N}, taken with OGLE-III photometry \citep{2008AcA....58...69U}. The measurements of extinction in $V$ and $I$ cover $\gtrsim$ 90 deg$^{2}$ of the bulge, nearly the entire OGLE-III survey area. As the spatial coverage was vast in both longitude and latitude, and covered a dynamical range of reddening from $0.6 \lesssim E(V-I) \lesssim 2.3$, there was enough information to disentangle the effects of extinction curve variations and the then-unknown geometrical configuration of the Galactic bar. Further, advances in filter technology allowed a calibration of the OGLE-III filters onto the system of \citet{1992AJ....104..340L} that was very accurate, eliminating what was previously a worrisome source of uncertainty. These maps are available on the OGLE webpage\footnote{\url{http://ogle.astrouw.edu.pl}}.

The extinction curve was confirmed to be unambiguously variable. Regression of the magnitude of the red clump versus its colour over scales as small as 30' yielded values spanning a range no smaller than $dA_{I}/dE(V-I) = 0.99 \pm 0.01$ up to $dA_{I}/dE(V-I) = 1.46 \pm 0.03$ (see Figure 7 of \citealt{2013ApJ...769...88N}). These variations can be found even at the same level of reddening, so they are not due to the convolution of the photometric bandpass with the extinction curve, or other non-linear effects. 

Demonstrating that the extinction curve is variable is not the same as solving for its variations. It was shown that the extinction curve could vary on scales smaller than 30', and thus the method of measuring regressions of extinction versus reddening toward relatively small regions was now demonstrated to be insufficient. \citet{2013ApJ...769...88N} discuss several failed attempts to constrain the total-to-selective extinction ratio as a function of sightline and why they did not work. 

\begin{figure}
\begin{center}
\includegraphics[totalheight=0.17\textheight]{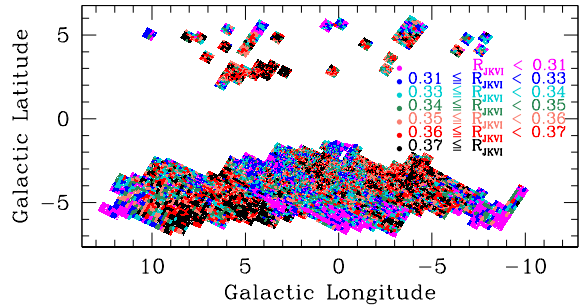}
\caption{This is Figure 12 of  \citet{2013ApJ...769...88N}. Shown is the ratio of $E(J-K_{s}$ measured by \citet{2012A&A...543A..13G} to $E(V-I)$ measured from OGLE photometry. The data is shown in equal area septiles. The extinction curve is clearly variable,  spanning the range $0.31 \rightarrow E(J-K_{s})/E(V-I) \rightarrow 0.17$ between the 14th and 86th percentiles.}
\label{RIJKMapper}
\end{center}
\end{figure}

What ultimately solved the issue was the combination of the $E(V-I)$ reddening maps from  \citet{2013ApJ...769...88N} with the $E(J-K_{s})$ reddening maps from \citet{2012A&A...543A..13G}, where the ratio is shown in Figure \ref{RIJKMapper}, which is Figure 12 of \citet{2013ApJ...769...88N}. The ratio of the two could be measured over scales as small as 3',  a correlation between $E(J-K_{s})/E(V-I)$ and $A_{I}/E(V-I)$ was measured and used to report the extinction everywhere. The best fit was found to be:
\begin{equation}
\begin{split}
A_{I}  &= 0.7465 \times E(V - I) + 1.3700 \times E(J - K_{s} ) \\
 & = 1.217\times E(V - I ){\times} \\
  & (1 + 1.126 \times(E(J - K_{s} )/E(V - I ) - 0.3433)).
\end{split}
\label{EQ:TwoColour}
\end{equation}
Assuming that $A_{I}/E(V-I)$ correlates with $E(J-K_{s})/E(V-I)$ is equivalent to assuming a single, dominant parameter to extinction curve variations (e.g. $R_{V}$). The precision in $A_{I}$ is estimated as $\sim$0.04 mag, and thus better than 4\%. This two-colour extinction correction was confirmed as being more reliable than any dereddening method based on $(V-I)$ colour alone in the analysis of bulge RR Lyrae stars by \citep{2015ApJ...811..113P}. It was shown that Equation \ref{EQ:TwoColour} gives a more reasonable and a substantially tighter distance distribution function to bulge RR Lyrae stars, and has thus been adopted by other RR Lyrae bulge studies \citep{2015ApJ...808L..12K}. 

\begin{figure}
\begin{center}
\includegraphics[totalheight=0.17\textheight]{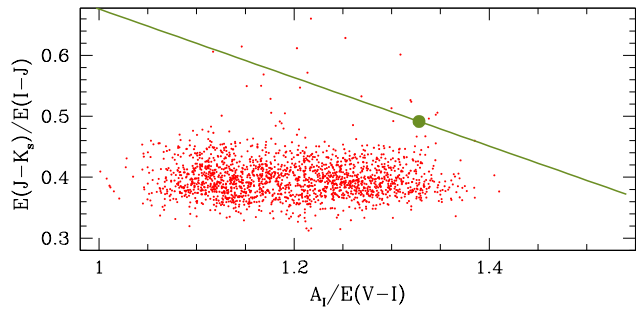}
\caption{This is Figure 8 of \citet{2016MNRAS.456.2692N}. The distribution of two extinction curve ratios, $E(J-K_{s})/E(I-J)$ and $A_{I}/E(V-I)$ is shown as the red points, with the prediction of \citet{1999PASP..111...63F}  shown in green with the green circle denoting the $R_{V}=3.1$ case. The predictions are poor match to the data regardless of the value of $R_{V}$. }
\label{ReddeningCoefficients2}
\end{center}
\end{figure}

The issue of extinction curve variations toward the bulge was followed up by \citet{2016MNRAS.456.2692N}, who added measurements of $E(I-J)$ to increase the diagnostic power -- three colours were measured rather than two. A comparison of two independent extinction curve ratios is shown in Figure \ref{ReddeningCoefficients2}. What \citet{2016MNRAS.456.2692N} demonstrated, above and beyond the previous findings of variable extinction, is that the extinction curve toward many bulge sightlines cannot be matched by the predictions of \citet{1989ApJ...345..245C} and  \citet{1999PASP..111...63F} \textit{regardless} of how $R_{V}$ is varied, demonstrating a total failure of those empirical frameworks toward the inner Milky Way. That is shown in Figure  \ref{ReddeningCoefficients2}, where not only does the prediction of \citet{1999PASP..111...63F} for $R_{V}=3.1$ (the green circle) fail to intersect the measurements, but rather the green line (variations in $R_{V}$) \textit{never} intersects the measurements. Further, as can also be seen in Figure \ref{ReddeningCoefficients2}, the variations in the extinction curve are not necessarily correlated between different extinction curve ratios. Both $A_{I}/E(V-I)$ and $E(J-K_{s})/E(I-J)$ vary significantly, but they do so independently. This necessitates additional degrees of freedom beyond $R_{V}$. 

\citet{2016MNRAS.456.2692N} then took the issue a step further and argued that interstellar extinction curve variations toward the bulge, which are at very high signal-to-noise and offer a large sample size due to the high reddening and high surface density of standard crayons, might be used as an essential laboratory for cosmology. The first motivation mentioned was the challenge of inferring dust properties toward Type Ia SNe. The mean extinction curve toward Type Ia SNe is $R_{V} \approx 2.05$ \citep{2015ApJ...813..137R}. The uncertainties in the treatment of extinction have been evaluated as the second largest source of systematic error in the determination of the dark energy equation-of-state parameter ``w" \citep{2014ApJ...795...45S}. This estimate assumes that the uncertainty is an uncertainty in $R_{V}$, and does not include the fact that for some sightlines the extinction curve is not fit by any value of $R_{V}$, which is now demonstrated as a fact of nature. 

\section{Extinction Toward Galactic Bulge Globular Clusters}
\label{sec:Globulars}
The list of $E(B-V)$ values toward bulge globular clusters to be found in the catalogue of \citet{1996AJ....112.1487H} would be difficult to make suitable in the context of this review. The literature sources are heterogeneous in methodology, accuracy, and precision. Further, the number reported in the catalogue, $E(B-V)$ is rarely or never the number actually measured by any of these methods for the heavily reddened bulge globular clusters. Rather, some other reddening index is measured, and then converted to $E(B-V)$ using reddening coefficients that are now demonstrated to be of dubious merit. 

One exception is the compilation of \citet{2005A&A...432..851R}, which was derived from the HST treasury program of \citet{2002A&A...391..945P}. Reddening determinations are made in the system of the observations, $E(F439W-F555W)$, by measuring the colour excess of the horizontal branch. In Table \ref{table:2}, we list the measurements of \citet{2005A&A...432..851R}, along with those of \citet{2012A&A...543A..13G} toward the same sightlines, using a $6'$ window input into the BEAM calculator.  From Table 11 and 12 of \citet{1995PASP..107.1065H}, we know that $E(F439W-F555W) \approx  E(B-V)$, not surprising as these are nearly identical filter pairs. 

The globular cluster NGC 6544 is an outlier, with $E(F439W-F555W)/E(J-K_{s})=1.07$. This is likely due to the globular cluster being relatively closeby, $\sim$3 kpc from the Sun \citep{1996AJ....112.1487H}, and thus only $\sim$120 pc below the Galactic plane and likely in front of a lot of the dust. For the remaining globular clusters, I measure $E(J-K_{s})/E(F439W-F555W)=0.41$. The prediction from \citet{1989ApJ...345..245C} is $E(J-K_{s})/E(F439W-F555W) \approx 0.53$.

\begin{table}
\centering
\begin{tabular}{|l|cc|}
	\hline \hline
Cluster Name & $E(F439W-F555W)$ & $E(J-K_{s})$ \\
	\hline \hline \hline
NGC 6380 & 1.58    &  0.54 \\
NGC 6401  & 0.90    &  0.45 \\
NGC 6453 & 0.61    &  0.28 \\
NGC 6522   & 0.53   &  0.24 \\
NGC 6540   & 0.52     &  0.20 \\
NGC 6544   & 0.76    &  0.71 \\
NGC 6569   & 0.56   &   0.21 \\
NGC 6638   & 0.38    &  0.17 \\
NGC 6642   & 0.43    &  0.16 \\
\hline
\end{tabular}
\caption{Compilation of reddening measurements toward bulge globulars from  \citet{2005A&A...432..851R} and toward the same sightlines by  \citet{2012A&A...543A..13G}.}
\label{table:2}
\end{table}

The high reddening toward bulge globular clusters has meant that they tend to be less studied than most other globular clusters, due to the greater difficulty of obtaining deep photometry. However, this is beginning to change, as these clusters are interesting in their own right. It is plausible that in the near future globular clusters may reprise their historical role as leading diagnostics of the extinction toward the bulge, given their potential for high-resolution, multi-wavelength extinction maps. For example, \citet{2012ApJ...755L..32M} measure differential reddening exceeding ${\delta}E(J-K_{s}) = 0.30$ over scales as small as 2" toward Terzan 5 \citep{2012ApJ...755L..32M}. Interested readers are referred to the review of bulge globular clusters found elsewhere in this special issue, \citet{2015arXiv151007834B}, for further information on systems.

%{2010A&A...515A..49R} 

\section{Discussion and Conclusion}
\label{sec:Conclusion}

The magnitude of progress in recent decades in the study of extinction toward the Galactic bulge is clearly high. The community has gone from its first extinction map covering an area of 40'$\times$40' near Baade's window \citep{1996ApJ...460L..37S}, to reddening maps spanning nearly the entirety of the bulge \citep{2012A&A...543A..13G}. It has shifted from assuming literature values of the interstellar extinction coefficients, to actively measuring them and in fact finding a $\sim$20-25\% offset \citep{2009ApJ...696.1407N,2016MNRAS.456.2692N}.

In spite of this progress, there remains a need for further progress, as the advances in this field are mirrored by advances in other areas of astronomy which necessitates superior accuracy and precision. Three areas in need of improvement are those of differential reddening estimates, understanding of the ``anomalous" extinction coefficients, and an integrated study of the planetary nebulae. 

Though the extinction maps of \citet{2012A&A...543A..13G} yield precise and accurate estimates of $E(J-K_{s})$ nearly everywhere toward the bulge, it's the case that differential reddening can exceed $\sim$10\% of the mean reddening toward bulge observing windows as small as 3' \citep{2013ApJ...769...88N}. This was estimated by measuring that the width of the red giant branch in colour space correlates with the mean reddening, which cannot be due to intrinsic factors as the gradient in metallicity \textit{dispersion} of the bulge is null or shallow \citep{2008A&A...486..177Z}. \citet{2012ApJ...755L..32M} measured differential extinction toward the globular cluster Terzan 5 exceeding $\sim$25\%, over a small field of $ 200" \times 200"$. Some progress will be needed on this effect as Galactic astronomy transitions to being a precision science, there may be hope by combining additional photometry from new surveys like the Blanco DECam Bulge Survey \citep{2014AAS...22334619C}. 

The issue of different extinction coefficients toward the bulge is also a strange one. I am aware of no theoretical prediction within the literature that would explain why this is so.  \citet{2014A&A...571A..91M} conjectures a ``Great Dark Lane'' between the Sun and the bulge, this could be the explanation, but at this time there have been no follow-up publications, and thus no estimates of the extinction curve specific to the Great Dark Lane. It was conceivable that the anomalous extinction could just be an artifact of incorrect assumptions as to what the ``standard" extinction curve is, but this conjecture has become less and less plausible. \citet{2003ApJ...590..284U} showed that the reddening toward the bulge is systematically different to that toward the Large Magellanic Cloud in a manner independent of systematics, by means of a purely differential analysis. \citet{2016MNRAS.456.2692N} combined four measures of extinction in the bandpasses $VIJK_{s}$ to show no compatibility between bulge extinction coefficients and literature extinction coefficients even after allowing for the range of ``standard" extinction curves to be found in the literature. \citet{2016arXiv160203928S} have recently confirmed that literature values of the ``standard" extinction curve are in fact not correct descriptions of nature even in the solar neighbourhood.  The discrepancy remains regardless, their Table 5 shows that their measured mean extinction coefficients for the local interstellar medium remain distinct from the measured values toward the bulge, with an offset of $\sim 16$\% for $E(V-I)/E(J-K_{s})$.

The planetary nebulae measurements are their own diagnostic, with completely independent systematics. Are the offsets in the Balmer decrement measured by \citet{1992A&A...266..486S} and \citet{2004MNRAS.353..796R} toward bulge planetary nebulae really due to additional opacity in the radio continuum, as argued by \citet{2013A&A...550A..35P}? If the extinction coefficients are non-standard, are we measuring the same phenomenon as measured with red clump and RR Lyrae stars? If they are in fact standard, how can that be reconciled with the measurement of extinction anomalies toward red clump and RR Lyrae stars? There is manifest potential for elucidation here, should there be an integrated study in the future.

\section*{Acknowledgments}
DMN was supported by the Australian Research Council grant FL110100012. I thank Albert Zijlstra for helpful discussions. I thank the anonymous referee for a constructive and detailed report.

\end{document}